\documentclass[aps,prl,twocolumn,showpacs,preprintnumbers,amsmath,amssymb]{revtex4}
\usepackage{epsfig}
\input epsf
\epsfclipon
\usepackage{epstopdf}

\begin{document}
\title{International trade network: fractal properties and globalization puzzle}
\author{Mariusz Karpiarz, Piotr Fronczak and Agata Fronczak}
\affiliation{Faculty of Physics, Warsaw University of Technology,
Koszykowa 75, PL-00-662 Warsaw, Poland}
\date{\today}

\begin{abstract}

Globalization is one of the central concepts of our age. The common perception of the process is that, due to declining communication and
transport costs, distance becomes less and less important. However, the distance coefficient in the gravity model of trade, which grows in time, indicates that the role of distance increases rather than decreases. This, in essence, captures the notion of the globalization puzzle. Here, we show that the fractality of the international trade system (ITS) provides a simple solution for the puzzle. We argue, that the distance coefficient corresponds to the fractal dimension of ITS. We provide two independent methods, box counting method and spatial choice model, which confirm this statement. Our results allow us to conclude that the previous approaches to solving the puzzle misinterpreted the meaning of the distance coefficient in the gravity model of trade.

\end{abstract}  \maketitle

The recent popular readings (e.g. \cite{bookCairncross,bookBarabasi}) provide ample anecdotal evidence for the so-called \emph{death of distance} and \emph{small world effect}. At the same time, the econometric society is rather restrained in advancing quantitative methods to study the globalization process. In particular, the econometric literature on the international trade fails to deliver consistent empirical support for globalization and for diminishing effects of distance on bilateral trade volumes. In this context, the most controversy is caused by the gravity model of trade, and the so-called \emph{distance} or \emph{globalization puzzle} that arises from the model.

\begin{figure}
\begin{center}
\centerline{\includegraphics[width=7cm]{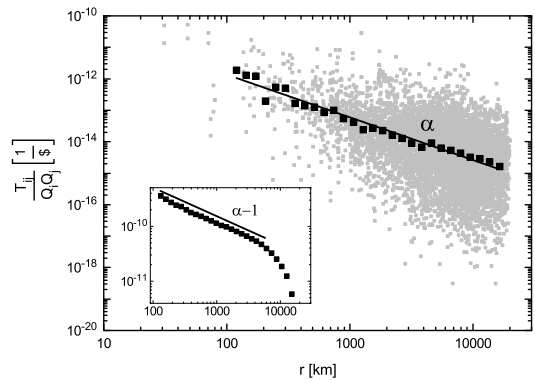}}
\caption{The effect of distance on trade. Main stage:
scattered plot of $v_{ij}\!=\!T_{ij}/(Q_iQ_j)$ versus $r_{ij}$, cf.~Eq.~(\ref{gravity}), for the year 1980.
Inset: cumulative measure $V(r)=\int_r^{\pi R}v_{ij}dr_{ij}$,
where $R$ is the Earth's radius. (The noise inherent to the data makes
difficult to clearly appreciate the power-law scaling of trade with distance. To overcome this problem and to estimate the value of $\alpha$, we have
defined the quantity, $V(r)$, which is integrated with respect to distance and corresponds to the area under the scattered plot shown in the main stage in the figure. The distance coefficient, $\alpha$, has been calculated from the slope of the linear part of the plot $\ln V(r)$ versus
$\ln r$, as it is shown in the inset. This procedure has been used to calculate distance coefficients for all the years in the period 1950-2000, see Fig.~\ref{fig2}).}\label{fig1}
\end{center}
\end{figure}

\begin{figure}
\begin{center}
\centerline{\includegraphics[width=8cm]{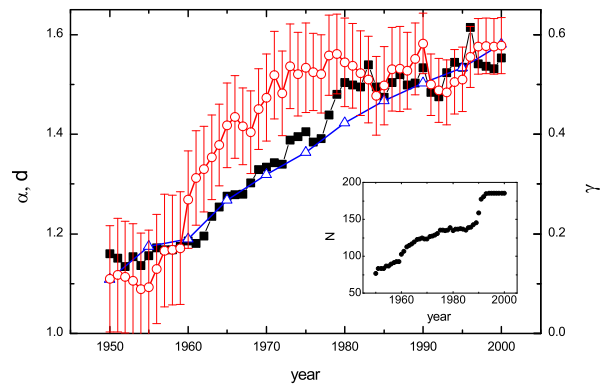}}
\caption{The year-by-year values of the distance coefficient, $\alpha$, (black squares) in comparison with: i. the fractal dimension $d$ obtained from the box counting method (open triangles), and ii. the parameter $\gamma$ obtained from the spatial choice model (open circles). Inset: the number of countries in the world.}\label{fig2}
\end{center}
\end{figure}

The gravity model of trade was first proposed in 1962 by Jan Tinbergen, the
physicist and the future first Nobel Prize Winner in Economic Sciences. Now,
the model is one of the most recognizable empirical models in economics
\cite{1998Deardorff,1979Anderson,1985Bergstrand,kaski,Fagiolo,Fronczak}. Drawing from Newton's law of gravity, the gravity model relates trade volume, $T_{ij}$, between two countries, $i$ and $j$, positively to the product of their GDP's, i.e. $Q_iQ_j$, and negatively to the geographic distance, $r_{ij}$, between them. The simplest form of the gravity equation for the bilateral trade volume is
\begin{equation}
T_{ij} = G\frac{Q_i Q_j}{r_{ij}^{\alpha}},
\label{gravity}
\end{equation}
where $\alpha$ is the distance coefficient, which is obtained from the real
data analysis (see Fig.~\ref{fig1}), and $G$ is a constant. The model,
Eq.~(\ref{gravity}), successfully explains trading patterns, but the growing
in time distance coefficient, $\alpha$, which is called the \textit{elasticity of trade with respect to distance} (see Fig.~\ref{fig2}), seems to indicate that the role of distance increases rather than decreases over time. In other words, distance appears more severe in spite of globalization. This, in essence, captures the notion of the missing globalization puzzle \cite{2002Coe,2005Brun,2008Disdier}.

Many explanations for this puzzle have been proposed in the literature,
starting from the continuously changing composition of trade
\cite{1999Rauch}, through the dispersion of economic mass across countries
\cite{1995Leamer}, ending at the introducing new quantities like multilateral resistance terms to the gravity equations \cite{2003Anderson}. However, because the proposed solutions usually lead to greater complexity of the original model, none of them has been accepted as fully satisfactory. In the following, we propose a completely new explanation for the distance puzzle. We argue that the coefficient $\alpha$ is strictly related to the fractal dimension of the international trade system (ITS), and changes in this parameter over time are due to spatio-temporal fractal evolution of ITS.

It is commonly accepted that the uneven spatial distribution of socio-economic activity can have fractal properties, that is, it can repeat itself at different levels of spatial aggregation \cite{2010Brakman}. As a prominent example, fractal organization of urban morphologies has been extensively explored \cite{Batty1994}. In most cases, the conjecture about the fractal character of studied objects (e.g. transport systems \cite{Holyst2005}, or wealth and population distribution \cite{Anazawa2004,Dauphin2012}) is usually based on scaling laws observed therein. However, it must be stressed that power laws do not necessarily certify the existence of a fractal structure. As an example may serve the famous Barabasi-Albert model for evolving networks \cite{Barabasi99}, in which a power law distribution of node degrees is not related to fractal properties. The same is true for simple maximum entropy network models of international trade (see e.g.~\cite{PRLGarlaschelli,Fronczak} which, although based on Pareto distributions for GDP, do not imply  fractality in the sense of self-similarity of complex networks \cite{NatureHavlin}. Therefore, our results concerning a spatio-temporal fractal nature of trade, which suggest that the globalization puzzle is rather related to changes in the spatial structure of the system, and not to trade-related costs, is a completely new contribution toward understanding the empirical gravity model of trade.

In this paper, we propose two different methods, both exploiting spatial properties of ITS, which allow to estimate its fractal dimension. The first one is a box counting method, which is a classical tool for such an analysis, and the second one, which exploits a simple decision-based model which is somehow related to the recently introduced radiation model for mobility and migration patterns \cite{Simini2012}. Having found the evidence of ITS fractality, we show that it can shed light on the origin of the globalization puzzle.

Our results are based on the trade data collected by K.S.~Gleditsch \cite{dataset} which contains, for each world country in the
period $1950-2000$, the detailed list of bilateral import and export volumes. The GDP data for all those years are taken from Penn-World Tables version 6.1 \cite{PWT}. The distance between countries is the distance between the countries' capitals, measured in kilometers \cite{distances}. All currency related
calculations were performed in U.S. dollars adjusted to the base year $1996$.

\begin{figure}
\begin{center}
\centerline{\includegraphics[width=\columnwidth]{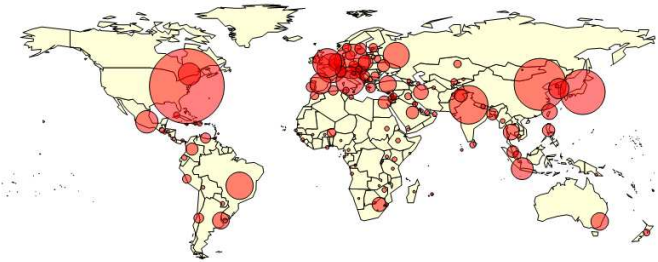}}
\caption{The object (i.e. set of circles of different radius) that reflects the topography of the real-world spatial economy in the year 1995. Description of the figure is given in the text.}\label{fig3}
\end{center}
\end{figure}

To begin, let us create an object that reflects the topography of the real-world spatial economy on which one can perform a simple box counting method to determine its fractal dimension. To this aim, each country is represented by the corresponding number of points, $S\propto GDP$, which are uniformly distributed inside a circle which is centered at the geographic center of the country. The basic idea behind this approach is that GDP of a country is, at first approximation, proportional to the number of places (i.e. points representing enterprises, companies, factories, etc.), in which GDP is produced. An example of such an object, in which area of each circle is proportional to $GDP$, is shown in Fig.~\ref{fig3} (see also supplementary material for a video demonstrating how this object changes in the consecutive years).

\begin{figure}
\begin{center}
\centerline{\includegraphics[width=7cm]{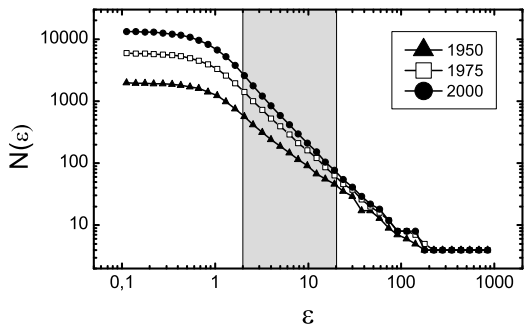}}
\caption{The number of nonempty boxes, $N$, needed to cover the object which is shown in Fig.~\ref{fig3} as a function of the size, $\varepsilon$, of boxes. The scale is expressed in steradians. The gray area represents the sizes from about $100$km to about $2000$km.}\label{fig4}
\end{center}
\end{figure}

Now, having such a well-defined object one can cover it with boxes of different sizes, $\varepsilon$, and find, how the number of nonempty boxes, $N$, changes with $\varepsilon$. (Fortunately, although the Mercator projection, which is used in this study, distorts the size and shape of large objects (especially near to the poles), all the countries are located in weakly distorted regions.) The results presented in Fig.~\ref{fig4} clearly show that there is a region (gray rectangle), where power law scaling allows one to estimate the fractal dimension of the object, $d=-\log N\!(\!\varepsilon\!)/\log \varepsilon$. In~Fig.~\ref{fig2}, the line with blue triangles represents results of the preformed analysis. The almost exact coincidence between the obtained fractal dimension, $d$, and the distance coefficient, $\alpha$, is remarkable. The above findings allows us to speculate that the distance coefficient, $\alpha$, is simply equal to the fractal dimension of the trade system.

\begin{figure}
\begin{center}
\centerline{\includegraphics[width=7cm,height=4cm]{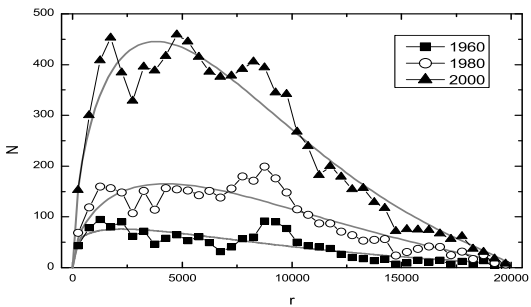}}
\caption{Histograms of the number of existing trade channels for three years: $1960$ (bottom), $1980$ (middle), and $2000$ (top). Solid lines represent the best fits of Eq.~(\ref{GammaSin}) to real data.}\label{fig5}
\end{center}
\end{figure}

Now, let us consider a spatial choice model which, on the one hand, describes the trade patterns of international trade (by the term \emph{trade pattern} we mean the number of trade connections, $T_{ij}>0$, vs their distance, as shown in~Fig.~\ref{fig5}), and, on the other hand, takes into account the fractal dimension of ITS. 

Spatial choice models are found in various choice contexts, where the agents (e.g. countries) are embedded in space and the pattern of connections (trade channels) they create depends on distance. For example, in nature, animals have to make decisions on how far to forage (optimizing trade-off between distance and amount of food) \cite{Sims2008}, in society, the people have to choose how far away from their work they should live
\cite{Expert_2011}.  In what follows, we adopt the model suggested in \cite{Malmberg_2011} to the international trade. 

\begin{figure}
\begin{center}
\centerline{\includegraphics[width=8cm]{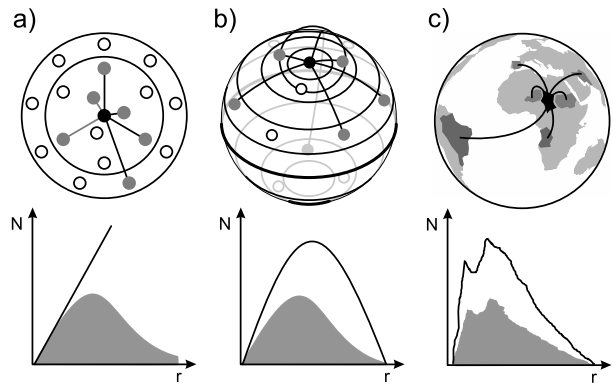}}
\caption{a) Basic spatial choice model in which agents are homogeneously distributed in the two-dimensional flat plane. b) Model with agents homogeneously distributed on the sphere. c) Model with agents heterogeneously distributed on the sphere. The corresponding diagrams below the schematic drawings show, how the number of available (black line) and realized (gray line with shaded area under it) connections depend on their length.} \label{fig6}
\end{center}
\end{figure}

According to the model, the empirical patterns of trade shown in Fig.~\ref{fig5} arise from two counteracting tendencies, both related with connection lengths. First, trade costs increase with distance, which provides an incentive to choose local (not too distant) trade partners. Second, the number of potential partners also depends on distance. In the case of a flat two-dimensional space, the number of homogeneously distributed
countries lying at a distance from $r$ to $r+dr$ from arbitrary country grows linearly with $r$, so the probability of finding a potential partner at $r$ increases with $r$ in the same way. Assuming that the cost related effect should finally overcome the effect of the increasing number of potential partners, one can show \cite{Malmberg_2011} that the continuous
choice process, which describes trade patterns, is given by the Gamma distribution
\begin{equation}
P(r)=\beta^{-2} r e^{-\frac{r}{\beta}}.
\label{Gamma1}
\end{equation}
In Eq. (\ref{Gamma1}) the factor $r$ corresponds to the linearly increasing
number of potential partners, while the exponential factor, $e^{-r/\beta}$, reflects decreasing choices due to the distance costs. The idea behind the model and the resulting trade pattern, Eq.~(\ref{Gamma1}), are shown in Fig.~\ref{fig6}a.

The above model can be significantly improved in order to better describe the empirical trade patterns. Firstly, it can be done by considering that the space in which the connections are formed is not flat. One could easily show that on the globe the number of potential trade partners (given their homogeneous distribution) increases with $r$ as $\sin(r/R)$, where $R$ is the Earth mean radius. It means that the flat Earth approximation, which is assumed in the basic spatial choice model, is appropriate for small distances not longer than several thousands of kilometers. And although for large distances, $r$, the exponential damping factor is sufficiently strong and growing with $r$ the number of potential partners has a minor significance, for intermediate distances trade patterns which result from flat and non-flat geometries may be quite different (see Fig.~\ref{fig6}b).

Secondly, as shown at the beginning of the paper, the assumption of uniform distribution of the countries is not entirely correct. To overcome this, one has to take into account the quasi-fractal structure of the continents and the area heterogeneity across the potential trade partners (see Fig.~\ref{fig6}c). It is reasonable to assume that the space in which trade takes place has a fractal dimension between $1$ (a line) and $2$ (a plane). Then, in the flat Earth approximation, the number of potential trade partners should depend on $r$ as $r^\gamma$, with $\gamma+1$ being the aforementioned fractal dimension. In this context, it is worth to remind that in general, the Gamma distribution, Eq.~(\ref{Gamma1}), belongs to a two-parameter family of continuous probability distributions of the form:
\begin{equation}
P(r)=\frac{\beta^{-\gamma-1}}{\Gamma(\gamma+1)}r^\gamma e^{-\frac{r}{\beta}},
\label{Gamma}
\end{equation}
where $\Gamma$ is the Gamma function. Note that setting in Eq.~(\ref{Gamma}) the parameter $\gamma$ as equal to $1$ one gets the ordinary Gamma distribution, Eq.~(\ref{Gamma1}), which corresponds to the case of the basic spatial choice model.

Finally, combining the Earth's sphericity with fractaltality one gets the following expression describing trade patterns:
\begin{equation}
P(r)=C \sin\left[(r/R_E)^\gamma\right] e^{-\frac{r}{\beta}},
\label{GammaSin}
\end{equation}
where $C$ is a normalization constant which, in opposite to the Gamma
distribution, has no simple analytic form.

Having the model defined, one can validate it by fitting Eq.~(\ref{GammaSin}) to the empirical trade patterns. In Fig.~\ref{fig5}, the solid gray lines represent the results of the fitting procedure for three different years. The fitting is very satisfactory. Moreover, the time dependence of the fitted parameter $\gamma(t)$, which is shown in Fig.~\ref{fig2} (red line with open circles), has the shape similar to both: the distance coefficient, $\alpha(t)$, in the gravity model of trade and the fractal dimension of ITS, $d(t)$. One can see that $\gamma(t)$ is shifted by a constant offset in comparison to $\alpha(t)$, i.e. $\gamma(t)=\alpha(t)-1$. This observation strongly supports our hypothesis that the parameter $\alpha$ in Eq.~(\ref{gravity}) has the meaning of the fractal dimension of ITS.

The hypothesis about the correspondence between the distance coefficient, $\alpha$, and the fractal dimension of the trade system can be heuristically justified by the fact that, in the relevant Newton's law, the corresponding coefficient results from the three-dimensional space in which the gravitational interaction is defined. In the Newton's law, however, the coefficient, $\alpha=2$, is equal to the dimension of the space, $d=3$, minus one, i.e. $\alpha=d-1$, whereas in the gravity model of trade the parameters are equal, $\alpha=d$. This may raise some suspicion as to the validity of the hypothesis. On the other hand, however, one should keep in mind that there is absolutely nothing fundamental in the formal analogy between the empirical laws of trade and gravity \cite{Squartini2013}. For the better analogy between the two, one can, for example, use the differential gravity law (the so-called \textit{tidal force}), instead of the Newton's law. It is remarkable that in the tidal force the distance dependence is characterized by $\alpha=d$. Furthermore, the force is used when the interacting objects are separated by distances that are not large compared to their physical size, what certainly takes place in the case of trading countries.

\begin{figure}
\begin{center}
\centerline{\includegraphics[width=7cm]{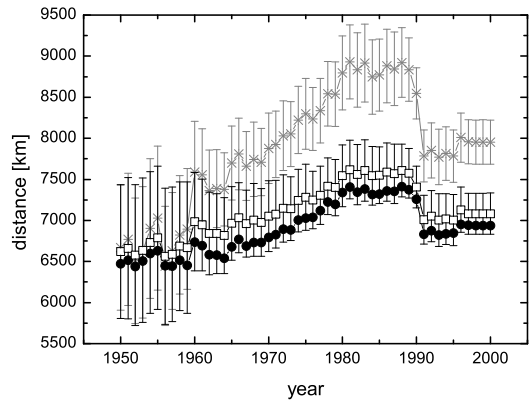}}
\caption{Time dependence of: i. the average length $\langle r\rangle$ of the trade connection (black circles), ii. the mean of the distribution given by Eq.~(\ref{Gamma1}) (gray stars), and iii. the mean of the distribution given by Eq.~(\ref{GammaSin}) (open squares).}\label{fig7}
\end{center}
\end{figure}

The relation between the coefficient $\alpha$ and fractality of the trade system, which is revealed in this paper, allows one to give a completely new solution to the globalization puzzle. Namely, it appears that the growth of $\alpha$ in time does not contradict the progress of globalization. Rather, it is a natural consequence of the growing density of trade connections which are embedded in a limited fractal-like space. This allows one to conclude that increasing in time character of the distance coefficient, $\alpha$, should not be associated with the growing role of distance or distance-related costs.
It can be seen by studying the average length of trade connections, which, as a rule, increases in time (see Fig.~\ref{fig7}). In that context, all previous approaches to solving the globalization puzzle were incorrectly motivated and perhaps unnecessary biased by the distant coefficient.

The work has been supported from the National Science Centre in Poland (grant no. 2012/05/E/ST2/02300).

\bibliography{fronczak}
\end{document}